\documentclass[usenatbib,fleqn]{mnras}
\usepackage{newtxtext,newtxmath}
\usepackage[T1]{fontenc}
\usepackage{ae,aecompl}

\usepackage{amsmath}	
\usepackage{amssymb}	

\usepackage{ctable}
\usepackage{url}
\usepackage{xspace}
\usepackage[normalem]{ulem}
\usepackage{hyperref}
\usepackage[all]{hypcap}
\usepackage{graphicx}

\def\msun{{\rm\,M_\odot}}

\def\tcc{t_{\rm cc}}
\def\msun{{\rm\,M_\odot}}

\newcommand{\be}{\begin{equation}}
\newcommand{\ee}{\end{equation}}

\newcommand{\fgas}{$ f_{\rm gas}$}
\newcommand{\rd}{$\rm R^*_{d}$}
\newcommand{\re}{$\rm R^*_{e}$}
\newcommand{\Msat}{ M_{\rm sat}}

\def\h2{${\rm\,H_2}$}

\usepackage{color}

\title[Enhanced SFR in JFGs]{Explaining the enhanced star formation rate of Jellyfish galaxies in galaxy clusters}
\author[Safarzadeh \& Loeb]{Mohammadtaher Safarzadeh$^{1,2}$\thanks{E-mail: msafarzadeh@cfa.harvard.edu} and Abraham Loeb$^{1}$
\\\\
$^{1}$Harvard-Smithsonian Center for Astrophysics, 60 Garden St. Cambridge, MA, USA\\
$^{2}$School of Earth and Space Exploration, Arizona State University, AZ, USA\\
}

\usepackage[all]{hypcap}

\pubyear{2019}

\begin{document}
\label{firstpage}
\pagerange{\pageref{firstpage}--\pageref{lastpage}}
\maketitle

\begin{abstract}

We study the recently observed JellyFish galaxies (JFGs), which are found to have their gas content ram-pressure stripped away in galaxy clusters.
These galaxies are observed to have an enhanced star formation rate of about 0.2 dex compared with a control sample of the same stellar mass in their disks.   
We model the increase in the star formation efficiency as a function of interacluster medium pressure and parametrize the cold gas content of the galaxies as a
function of cluster-centric distance. We show that regarding the external pressure as a positive feedback results in agreement with the observed distribution of enhanced 
star formation in the JFGs if clouds are shielded from evaporation by magnetic fields. Our results predict that satellites with halo mass $< 10^{11}\msun$ moving with Mach numbers $\mathcal{M}\approx2$, and inclination angles below 60 degrees, are more likely to be detected as JFGs.  

\end{abstract}


\section{Introduction}

Ram pressure stripping (RPS) of the gas content of the galaxies entering a cluster-like environment  \citep{Gunn:1972gx} have long been observed \citep{Kenney:2004fn,Chung:2007im,Boselli:2014cn,Brown:2017kf,Hayashi:2017cx}, and 
modeled in numerical simulations \citep{Mayer:2006ee,Roediger:2007kr,Tonnesen:2009co,Steinhauser:2016da,Ruggiero:2017kc,Quilis:2017bc,Safarzadeh:2017jo,Yun:2018ja}. 
However, depending on the density and size of the stripped gas clouds, these clouds can survive for sufficiently long time in the cluster environment to form stars. 

Extreme examples of such galaxies have been observed and named Jellyfish galaxies, hereafter JFGs \citep{Poggianti:2016el}. 
These galaxies are found in all clusters and at all cluster-centric distances. JFGs show an enhanced level of star formation both in their disk and their tails \citep{Vulcani:2018bb},
that depends on how gas-rich the satellite is when it enters the cluster, and on how fast the gas is being stripped away. 

The enhancement of the star formation under the external pressure in galaxy clusters has been studied previously.
\citet{RamosMartinez:2018iv} showed that in interaction of a face-on disk with the interacluster medium (ICM), gas from the outskirts of the disk flows towards the center leading to high gas surface densities  and therefore higher 
star formation rate. \citet{Bekki:2010fl} showed that the strong compression of cold gas in gas-rich cluster members leads to a starburst phase of the satellites when the ICM pressure is boosted in cluster merging events. 
\citet{Bieri:2015eb} showed that AGN feedback can act as a positive feedback by over-pressurizing the star forming regions of a galaxy and therefore leading to a higher star formation rate (SFR). 
\citet{Bekki:2013hn} showed that the impact of ram pressure on the star formation activity of a satellite would depend on the pericenter distances of its orbit, the inclination angles of the disk, and the halo mass
dependency. \citet{Bekki:2013hn} show that after pericentre passage, the SFR could increase or decrease depending on the satellite's halo mass. \citet{Kapferer:2009fu} found that the enhanced SFR due to ram pressure stripping shows up in the wake 
of the galaxy rather than the disk and should be a more sensitive function of the ambient density than the relative velocity. On the other hand, \citet{Tonnesen:2009co} have found that only low ram pressures can compress
the gas into high density clouds and therefore lead to higher SFR, while high ram pressure will likely strip the gas rather than compress it. 

Here we focus on the timespan during which the cold gas is being stripped and show how to calculate the elevated SFR in the satellite prior to full stripping of the gas \citep{Rafieferantsoa:2018wv}.
We assume magnetic insulation of the molecular clouds in the disk from the hot ICM as inferred in cold fronts within the ICM \citep{Vikhlinin:2001bx}.

The structure of the paper is as follows. In \S (2) we show how ram pressure stripping can act as a positive feedback mechanism for increasing the star formation efficiency in the 
satellites' disk. In \S (3) we show how integrating this positive feedback with the giant molecular cloud mass function can lead to the observed increase in the SFR of JFGs, and 
in \S (4) we discuss the implications and predictions of our model.
 
\section{Ram pressure as a positive feedback}

The SFR of a galaxy is closely related to its gas surface density \citep{Kennicutt:1998id}. 
The gas content of the satellites depends on their halo mass and redshift. Main sequence galaxies are observed to be more gas rich at higher redshifts \citep{Daddi:2010ee,Tacconi:2010cr} with large cold gas fractions, \fgas $\equiv \rm \frac{M_{ gas}}{M_{gas}+M_{star}}$ \citep{Geach:2011fr,Narayanan:2012bd,Popping:2012iv,MorokumaMatsui:2015gk,Popping:2015fs}. 

In order to construct a stellar and gas disk of the satellites, we assume the cold gas is in the form of an extended exponential disk profile with gas scale radius $R_d^g=1.7 R^*_d$ \citep{Popping:2015fs}, 
and a stellar radius, $R^*_d$ computed as \rd=\re/1.67, where \re~is the half-light radius of the stellar disk. 
\re~is related to halo mass of the galaxy as, \re=0.015 $R_{200}(\Msat)$ \citep{Kravtsov:2013cy}, where 
the halo radius ($R_{200}$) is defined as enclosing an overdensity of 200 times critical density of the universe at a
given redshift. Therefore, knowing the halo mass of a galaxy, one can estimate the cold 
gas scale length which would be related to the maximum size of the cold clouds that would be stripped from a satellite. We estimate the stellar disk mass ($M_*)$ given a halo's redshift and mass following \citet{Behroozi:2013fga}.

When a satellite galaxy enters the hot ICM, it encounters the thermal and ram pressure of the ICM given by:
\be
P_{tot} = \rho_{ICM} c_s^2 (1+\mathcal{M}^2) \cos (\theta)
\ee 
where $\rho_{ICM}$ is the ICM density, $c_s$ is the sound speed of the ICM, and $\mathcal{M}$ is the Mach number of the satellite moving in the ICM.
We denote the a angle between the angular momentum vector of the satellite's disk and its direction of motion in the cluster as $\theta$, such that $\cos(\theta)=1$ for a 
face-on configuration.
We assume a fiducial Mach number of $\mathcal{M}=1.4$ for the satellites \citep{Faltenbacher:2005dy}, and we consider a maximum Mach number of $\mathcal{M}=2$ corresponding 
to those satellites moving with approximately escape velocity of the cluster. 
The external pressure acts as a positive feedback and triggers star formation in the disk \citep{Swinbank:2011df}. The resulting SFR depends on the 
ratio of the external pressure and the ISM pressure of the satellite's disk \citep{Elmegreen:1997he}.
The ISM pressure of the satellite before infall is given by:

\be
P_{ISM}\approx \frac{\pi}{2} G \Sigma_{g} \bigg[\Sigma_{g}+(\frac{\sigma_g}{\sigma_*})\Sigma_*\bigg],
\ee
where we model both stellar and gas surface densities as exponential profiles and assume $\sigma_g\approx\sigma_*$. Figure 1 shows the ISM pressure for satellites of different halo mass pre-infall. 
\begin{figure}
\resizebox{3.5in}{!}{\includegraphics{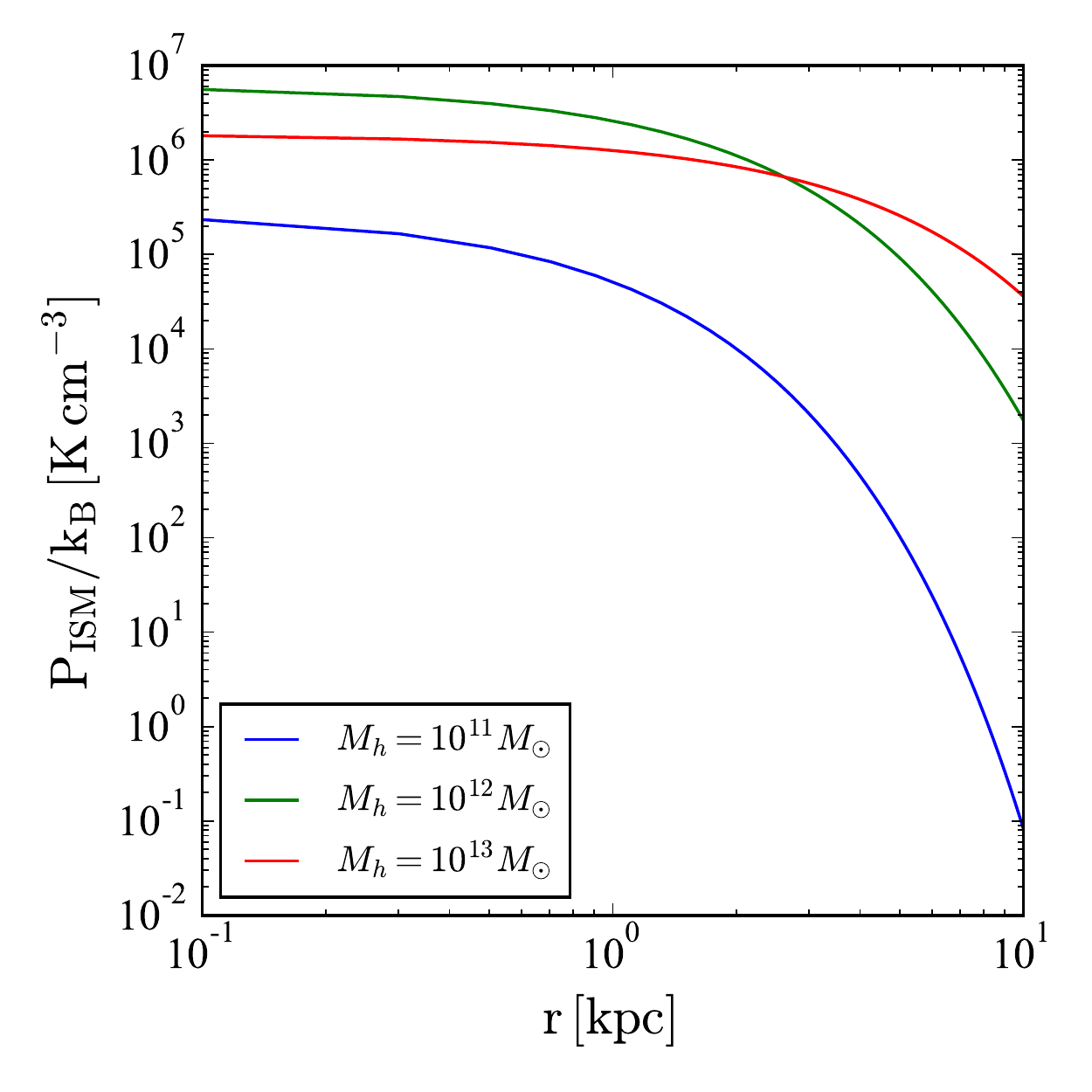}}
\caption{ The ISM pressure of galaxies with different halo masses pre-infall. The gas fractions are assigned given the stellar mass of the 
galaxy and its redshift following \citet{Popping:2015fs}. The stellar masses are assigned based on \citet{Behroozi:2013fga}. }
\label{f.cluster}
\end{figure}

\begin{figure}
\resizebox{3.5in}{!}{\includegraphics{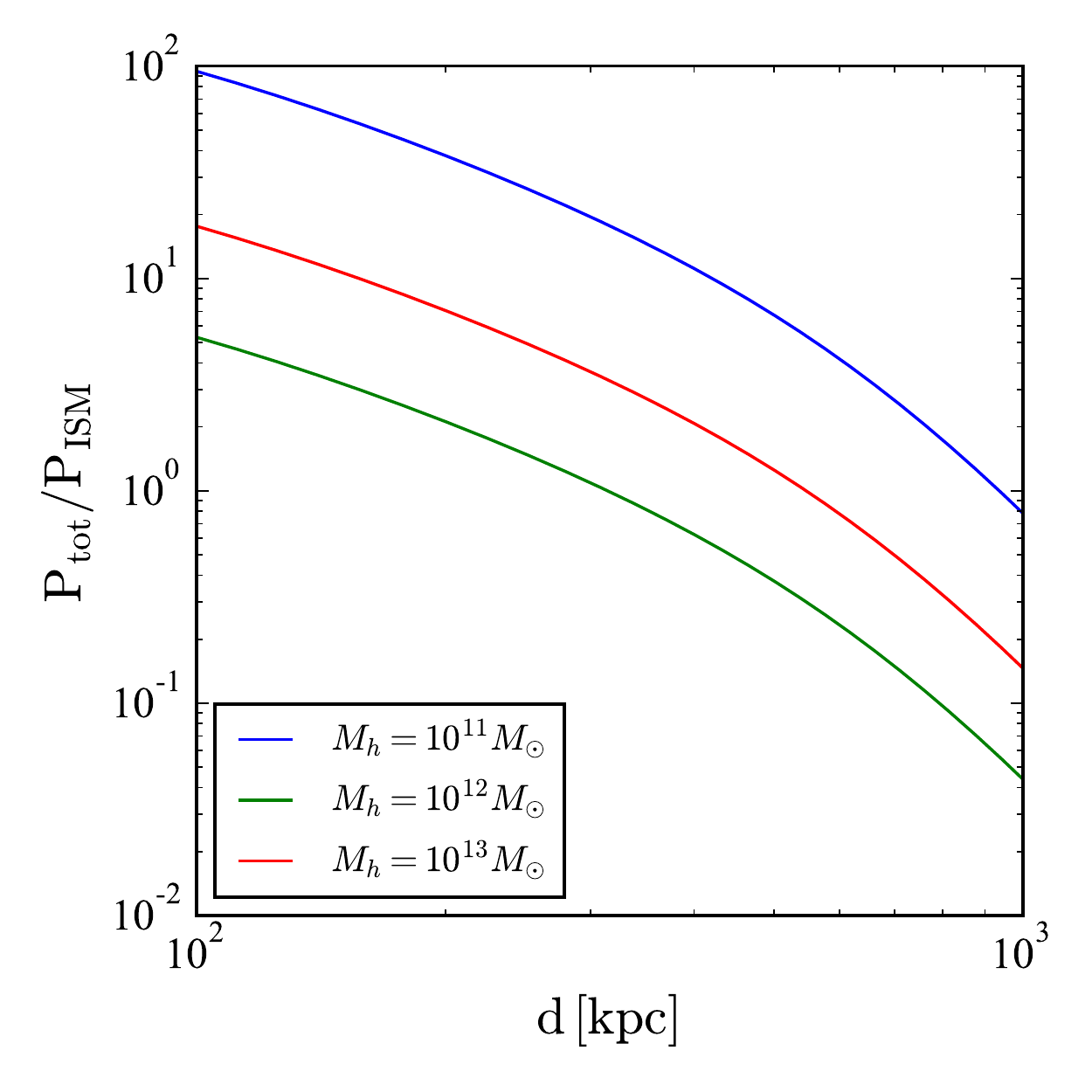}}
\caption{ The ratio of the ICM pressure to the ISM pressure of the disk of the satellites at the location of the effective radius of the disk. 
The gas density and temperature profile of A1795 is modeled following \citet{Vikhlinin:2006fp}. 
We have assumed $\mathcal{M}=1.4$ for the satellites.}
\label{f.cluster}
\end{figure}

\begin{figure*}
\resizebox{3.in}{!}{\includegraphics{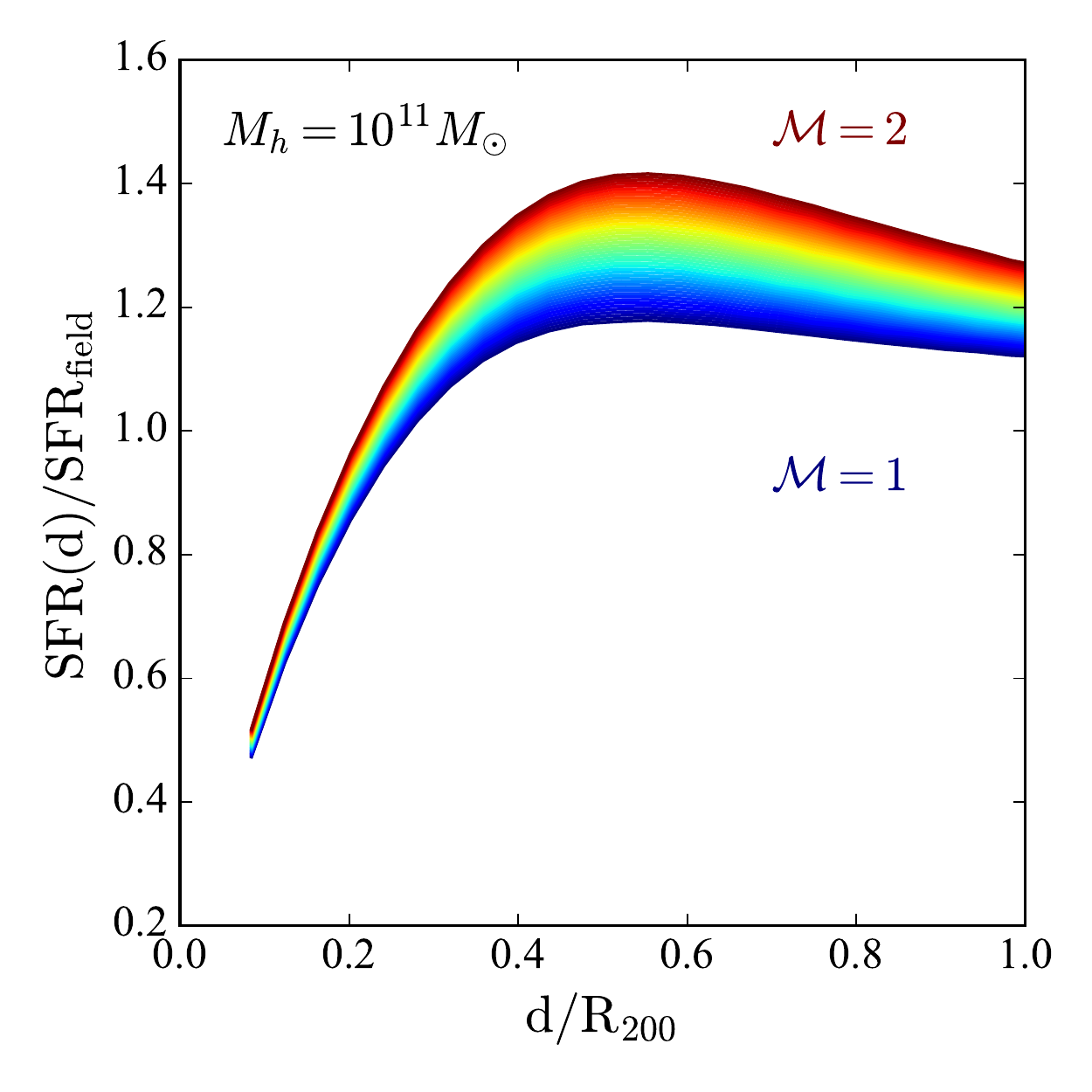}}
\resizebox{3.in}{!}{\includegraphics{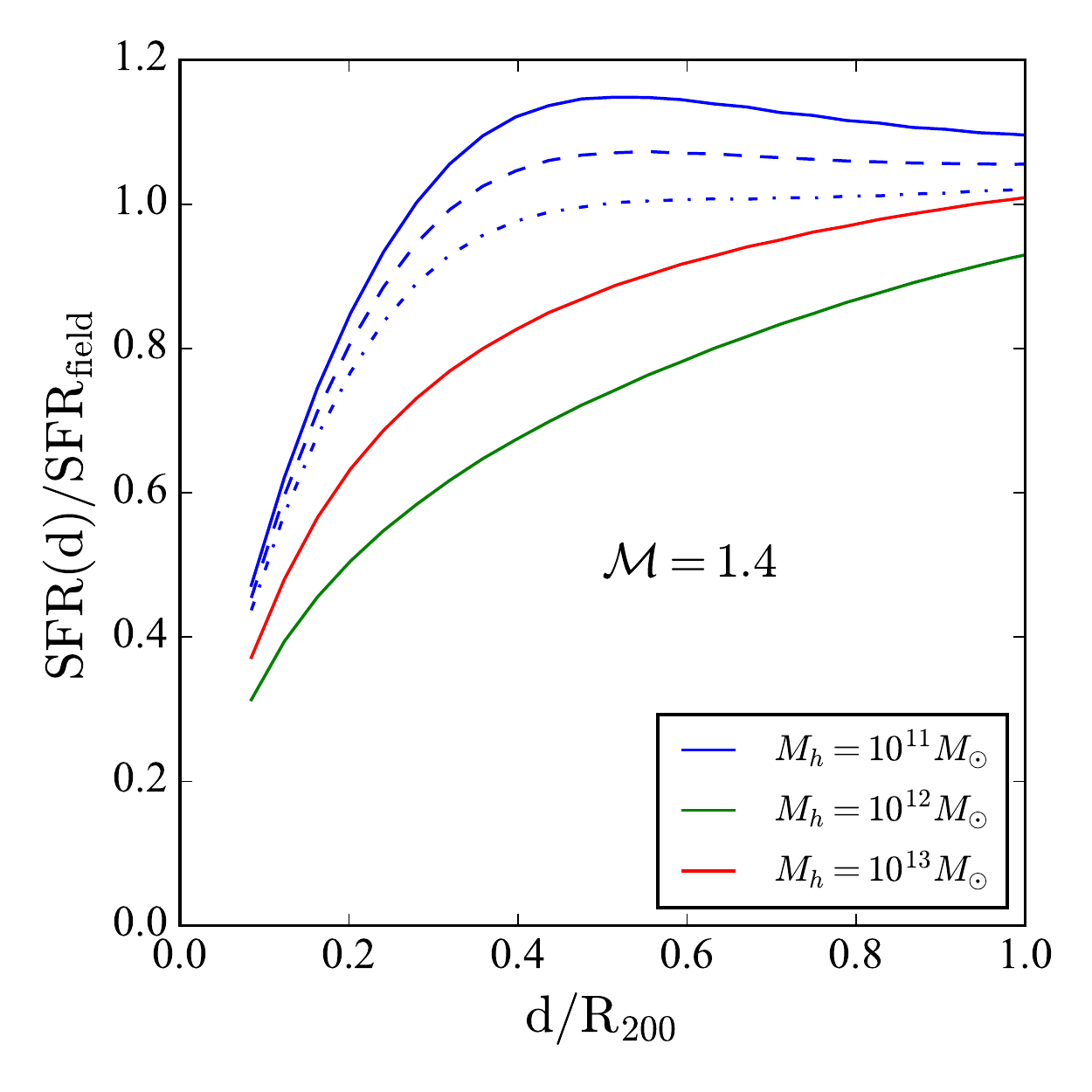}}
\caption{{\it Left panel}: the ratio of the SFR of the satellites as a function of 
cluster-centric distance to the SFR of the satellite in the field, shown for a satellite with halo masses of $10^{11}\msun$ moving in a host galaxy cluster of mass $\sim2\times10^{14}\msun$. Lines are color coded from lowest to the highest corresponding to Mach numbers values ranging from $\mathcal{M}=1$ to $\mathcal{M}=2$ which is about maximum seen for satellites in galaxy cluster simulations \citep{Faltenbacher:2005dy}. We are showing the enhancement of the integrated SFR over the entire disk. 
{\it Right panel}: results for different satellite halo masses assuming the satellite is moving with $\mathcal{M}=1.4$. The blue dashed and dot-dashed lines show the case where the inclination angle of the satellite is assumed to be
45, and 60 degrees, respectively. At inclinations above 60 degrees we expect the enhancement of the SFR in the disk to be significantly suppressed.}
\label{f.cluster}
\end{figure*}

\begin{figure}
\resizebox{3.5in}{!}{\includegraphics{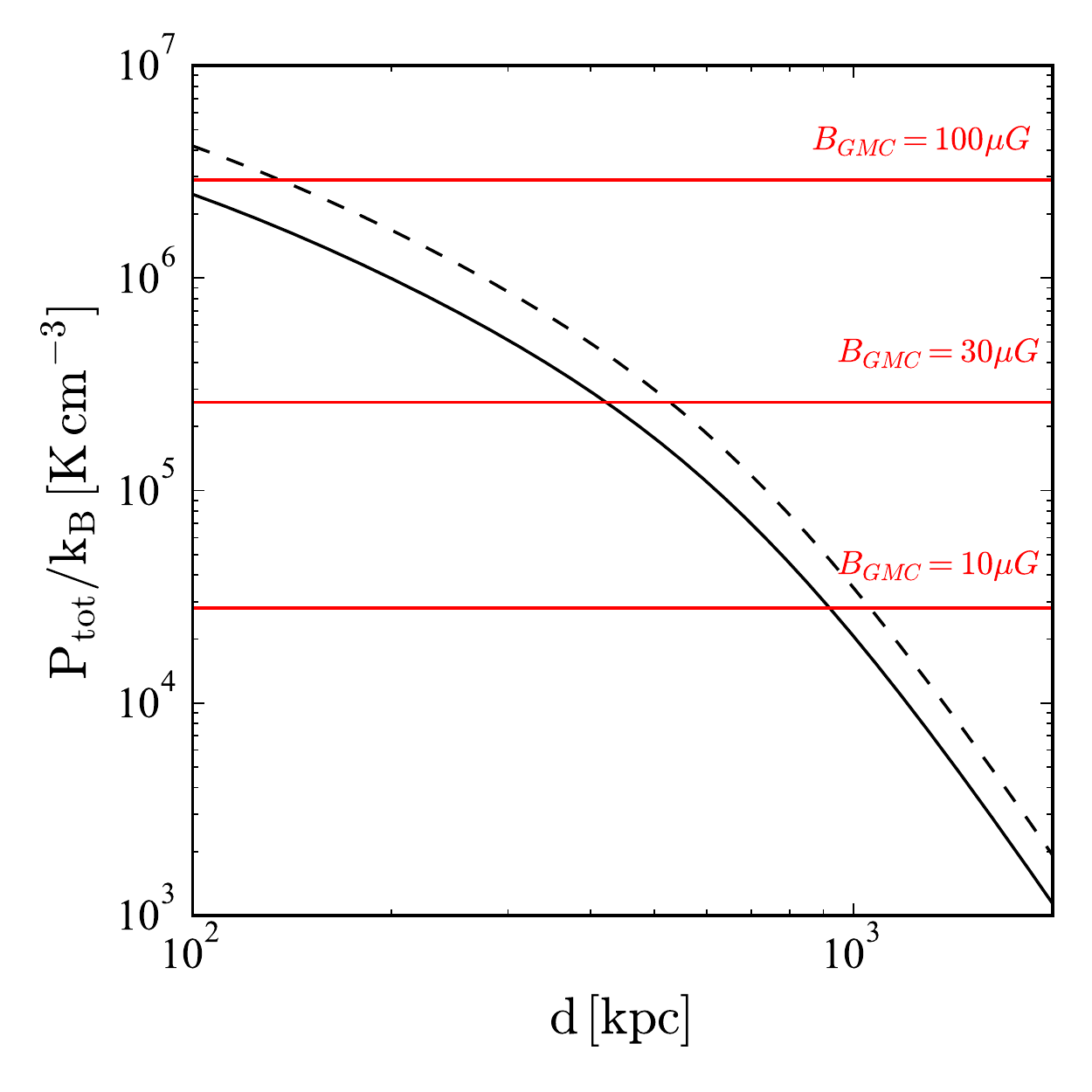}}
\caption{The ICM pressure for a satellite moving with $\mathcal{M}=$ 1.4 (2) is shown in solid (dashed) black lines respectively. The red horizontal lines indicate the magnetic 
field energy density in the GMCs for different assume values as labeled close to each line. GMC magnetic field strength of 30 $\rm \mu G$ can shield the molecular clouds down to about 400 kpc in a host cluster of mass $\sim2\times10^{14}\msun$ given the Mach number of the satellite.}
\label{f.cluster}
\end{figure}

To compute the ram pressure that the satellite encounters, we model our galaxy cluster with gas density profile similar to A1795 as $\sim$ 10 of the known population of the JFGs have been detected in this galaxy cluster \citep{Poggianti:2016el}. 
The gas density and temperature profile of A1795 is modeled following \citet{Vikhlinin:2006fp}.
We compute the ram pressure assuming that the satellite moves in the ICM with Mach number $\mathcal{M}\approx1.4$. 

The ratio of the ICM pressure to the ISM pressure of the disk of the satellites at the location of the effective radius of the disk is shown in Figure 2. 
The regime where the pressure ratio is greater than 1, we expect positive feedback to enhance the SFR in the satellites' disk. 

\section{Enhanced Star formation efficiency}

The impact of the external pressure on the star formation efficiency has been formulated in \citet{Elmegreen:1997he}. High pressures increase the star formation efficiency more 
effectively for more massive clouds.  In our case, the increase in the pressure is modeled as $P_{tot}/P_{ISM}$. 
In order to get a sense of what types of clouds contribute to the enhancement of the star formation in JFGs, we would need to specify the 
giant molecular cloud (GMC) mass function. 

We assume the GMC mass function of $dN/dM \propto M^{-\alpha}$ with $\alpha=-2$ \citep{Hopkins:2012hg}. 
The mass weighted star formation efficiency at a given pressure is computed as, 
\be
\epsilon (P) = \frac{\int^{M_u}_{M_l} \epsilon (M,P) \frac{dN}{d ln M} dM}{\int^{M_u}_{M_l}  \frac{dN}{d ln M} dM}, 
\ee
where $\epsilon (M,P)$ is computed following \citet{Elmegreen:1997he}.
We only consider clouds in the mass range of $M_l=10^4\msun$ to $M_u=10^6\msun$ to be relevant for our study, where 
the upper limit is based on the observations of the GMCs in the Milky Way galaxy \citep{Rosolowsky:2005gt,Fukui:2010ki}. 
Pressure is related to the cluster-centric distance ($d$), and therefore one can express $\epsilon (P) $ as $\epsilon (d) $. 

Since the observations trace the SFR, one has to connect the enhanced star formation efficiency to SFR. We write $SFR \propto \epsilon \times M_{\rm cold}$. 
Modeling the cold gas reservoir to decline with cluster-centric
distance provides a reasonable physical picture. We model the depletion of cold gas mass to be approximately a smooth linear function of the cluster-centric distance such that the cold gas mass at cluster-centric distance $d$ is $M_{\rm cold,d}/M_{\rm cold,field} \propto d/R_{200}$. Although more sophisticated treatment would be needed to understand how cold gas is removed from the galaxy by both stripping and star formation \citep{Zinger:2018io}, 
we proceed with this simple ansatz with two implications: (i) the lost cold gas mass reduces the internal ISM pressure of the satellite, and therefore enhances the relative pressure of the ICM to the ISM; and (ii) it directly influences 
the gas mass available for the star formation. 

In the left panel of Figure 3 we show the ratio of the SFR of a $10^{11}\msun$ satellite as a function of cluster-centric distance for different values of the Mach numbers. Lines are color-coded from the lowest to the highest
Mach number with dark blue to dark red. Right panel of Figure 3 shows the result for satellites of different halo mass moving with $\mathcal{M}=1.4$ in the ICM. 
A satellite with halo mass of $\approx 10^{11} \msun$ is more likely to be considered as a JFG. Halos with SFR enhancement of about 50\% are distributed over a wide cluster-centric distances 
but mostly found between $0.4 R_{200}$ and $0.8 R_{200}$ of their host cluster, in good agreement with the observations of the JFGs in OMEGAWINGS sample \citep{Poggianti:2016el}.
Moreover, we show the impact of the inclination angle $\theta$ on the same panel. The blue dashed and dot-dashed lines indicate a situation where the inclination angle is assumed to be 45 and 60 degrees, respectively, for a satellite 
with halo mass of $10^{11} \msun$. At inclinations above 60 degrees, we expect the impact of the ram pressure on enhancing the star formation rate of the galaxy to be significantly suppressed.

\section{Discussion}

\citet{Fujita:1999kr} studied the impact of ram pressure on the SFR of satellites in clusters and find a factor of 2 increase in SFR at pericenter distances. 
In their model the molecular gas is never stripped and the SFR rapidly drops near the central regions because the cold gas is stripped away and the existence of molecular clouds is truncated. 
Although the spirit of our method is similar to their approach, we treat differently the process and compare our results to new data on JFGs . 

\citet{Scannapieco:2015je} showed that in the absence of a shielding magnetic field layer, a cold cloud in the ICM would be disrupted on a timescale given by:
\be
t=\alpha t_{cc} \sqrt{1+\mathcal{M}};
\ee
with $t_{cc}$ being the cloud crushing timescale, defined as:
\be
t_{cc}\equiv\frac{\chi_0^{1/2} R_{\rm cloud}}{v},
\ee
where $v$ is the speed of the cloud in the ICM, $R_{\rm cloud}$ is the radius of the cloud, $\chi_0$ is the initial density ratio between the cloud and ICM, and 
$\alpha$ quantifying the fraction of the cloud that is disrupted after some given time. 
For the case of a cloud with mass of $10^5\msun$, and radius of 30 pc, the time it takes for the cloud to be 90\% disrupted moving with $\mathcal{M}=2$ in the ICM of A1795 ranges from about 
45 Myr at distances of 1 Mpc from the cluster center, to 3 Myr at a distance of 10 kpc. These timescales are too short with respect to the time it takes the satellite to reach the inner part of the 
galaxy cluster, meaning the clouds would have been largely disrupted by the ICM pressure.

The underlying assumption in our model is that the molecular clouds are shielded by magnetic fields from evaporation due to the hot ICM.
Magnetic fields in spiral galaxies have two separate elements: (i) the ordered magnetic field which is at the order of a few $\rm \mu G$, and (ii) the turbulent component that is an order of 
magnitude larger in amplitude \citep{Beck:2015fu}. Very strong magnetic fields of order of a few $\rm mG$ has been reported towards highly star forming regions \citep{Robishaw:2008bq,McBride:2013bc,Han:2017jo} 
through OH masers and measures of their Zeeman splitting. 

Figure 4 shows the comparison between the energy density in the magnetic field of the GMCs and the pressure encountered in the ICM.
Estimating the magnetic field energy density, $u_B=B^2/8\pi$ and equating it to the total pressure from ICM on a satellite 
moving with Mach number of $\mathcal{M}=1.4$ we find that a GMC magnetic field strength of 30 $\rm \mu G$ can shield the molecular clouds down to about 400 kpc in a host cluster of mass $\sim2\times10^{14}\msun$. This level of 
magnetic field is only expected in GMCs and is separate from the ordered magnetic field that is of the order of a few $\rm \mu G$. 

Our results are consistent with the observed distribution of JFGs in IllustrisTNG Gravity and MHD simulation \citep{Yun:2018ja} where the JFG candidates are shown to occupy the higher end of supersonic velocities
and reside in smaller satellite halos masses. In our analysis we have parametrized the ISM pressure of the satellites assuming they resemble disk galaxies in the local universe. The cold gas fraction of halos increase with redshift, 
which results in an increase in their ISM pressure. If we model the satellite to resemble those at $z~0.1$ before infall, we would not see an increase in the SFR in their disk since $P_{\rm tot}/P_{\rm ISM}$ is low for such systems.
Therefore, we require the JFGs to be late infallers for our model to work. Moreover, we predict no JFGs to be present at short cluster-centric distances, which is observed in simulations \citep{Yun:2018ja}  and observations \citep{Poggianti:2016el}.

\section{Acknowledgements}

We are thankful to the referee for their useful comments. 
We are also thankful to Daisuke Nagai, Evan Scannapieco, and Sownak Bose for helpful comments. This work was supported by the National Science Foundation under grant AST14-07835 and by NASA under theory grant NNX15AK82G as well as a JTF grant. MTS is grateful to the Harvard-Smithsonian Center for Astrophysics for hospitality during the course of this work. 

\bibliographystyle{mnras}
\bibliography{the_entire_lib}

\end{document}